\documentclass[aps,floatfix,prl,superscriptaddress,reprint,showpacs,10pt,preprintnumbers,longbibliography]{revtex4-2}
\usepackage[utf8]{inputenc}
\usepackage{lmodern}
\usepackage{amsmath,amssymb}
\usepackage[version=4]{mhchem}
\usepackage{dsfont}
\usepackage{mathtools}
\usepackage{graphicx}
\usepackage{url}
\usepackage{bbold}
\usepackage{color}
\usepackage{enumitem}
\usepackage[dvipsnames]{xcolor}
\usepackage[colorlinks=true,breaklinks=true]{hyperref}
\hypersetup{allcolors=[rgb]{0.0 0.0 0.6},linkcolor=[rgb]{0.75 0.05 0.05}}
\usepackage{bm}
\usepackage{epsfig}
\usepackage{amsmath}
\usepackage{amssymb}
\usepackage{slashed}
\usepackage{color}
\usepackage{accents}
\usepackage[dvipsnames]{xcolor}
\usepackage[colorlinks=true,breaklinks=true]{hyperref}
\usepackage[normalem]{ulem}
\hypersetup{allcolors=[rgb]{0.0 0.0 0.6},linkcolor=[rgb]{0.75 0.05 0.05}}
\DeclareMathOperator{\Tr}{Tr}

\newcommand{\be}{\begin{equation}}
\newcommand{\ee}{\end{equation}}
\newcommand{\ba}{\begin{eqnarray}}
\newcommand{\ea}{\end{eqnarray}}

\begin{document}

\title{Time- and Space-Varying Neutrino Mass Matrix from Soft Topological Defects}
\author{Gia Dvali}
\affiliation{
Arnold Sommerfeld Center, Ludwig-Maximilians-Universit\"at, Theresienstra{\ss}e 37, 80333 M\"unchen, Germany
}
 \affiliation{
Max-Planck-Institut f\"ur Physik, F\"ohringer Ring 6, 80805 M\"unchen, Germany
}
\author{Lena Funcke}
\affiliation{Center for Theoretical Physics and NSF AI Institute for Artificial Intelligence and Fundamental Interactions, Massachusetts Institute of Technology, 77 Massachusetts Avenue, Cambridge, MA 02139, USA}
\author{Tanmay Vachaspati}
\affiliation{Physics Department, Arizona State University, Tempe, AZ 85287, USA}

\begin{abstract}
We study the formation and evolution of topological defects that arise in the post-recombination phase transition predicted by the gravitational neutrino mass model in Ref.~\citep{Dvali2016a}. In the transition, global skyrmions, monopoles, strings, and domain walls form due to the spontaneous breaking of the neutrino flavor symmetry. These defects are unique in their softness and origin; as they appear at a very low energy scale, they only require Standard Model particle content, and they differ fundamentally depending on the Majorana or Dirac nature of the neutrinos. 
One of the observational signatures is the time dependence and space dependence of the neutrino mass matrix, which could be observable in future neutrino experiments. Already existing data rule out parts of the parameter space in the Majorana case. The detection of this effect could shed light onto the open question of the Dirac versus Majorana neutrino nature.
\end{abstract}

\maketitle

\textit{Introduction.}---The origin of the observed small neutrino masses is one of the greatest puzzles of the Standard Model (SM). The most popular directions of model building
beyond the SM usually focus on new physics at high-energy scales. Many high-energy origins of neutrino masses have been proposed, which require either hypothetical new particles~\citep{Minkowski1977,Mohapatra1979,Gell-Mann1979,Yanagida1980} or extra dimensions~\cite{ArkaniHamed1998,Dienes1998,Dvali1999} (see Ref.~\cite{Senjanovic2011} for a review). These scenarios all depend on electroweak symmetry breaking, and the resulting neutrino masses are proportional to (some power of) the SM Higgs vacuum expectation value.

As an alternative direction, a new low-energy solution to the neutrino mass problem was proposed in Ref.~\cite{Dvali2016a}, which is independent of the SM Higgs vacuum expectation value and does not require any additional dimensions or particles. Using the SM particle content coupled minimally to gravity, it only requires the assumption that pure gravity contains a non-zero topological vacuum susceptibility, 
\begin{align}\label{eq:RRcorr}
\langle R\tilde{R},R\tilde{R}\rangle_{q\to 0} &\equiv\lim\limits_{q \to 0}\int d^4 x\, e^{iqx} \langle T[R\tilde{R}(x)R\tilde{R}(0)]\rangle\nonumber\\&= \mathrm{const}\neq 0 ,
\end{align}
in the limit of vanishing momentum $q$. Here, $T$ is the time ordering operator, $R$ is the Riemann tensor, and $\tilde{R}$ is its dual. 
The condition [Eq.~\eqref{eq:RRcorr}] is equivalent to the statement that the gravitational analog of the QCD $\theta$-term,
\begin{equation}
\mathcal{L}_G\supset\theta_G R\tilde{R},
\label{eq:RRtheta} 
\end{equation}
is physical~\cite{Delbourgo1972,Eguchi1976, Deser1980,AlvarezGaume1984}, meaning that the free parameter 
$\theta_G \in [0,2\pi ]$ is a physically measurable quantity. 

If the condition [Eq.~\eqref{eq:RRcorr}] is fulfilled, it was shown that fermion condensation $\langle \bar{f}f\rangle$~\citep{Dvali2005,Dvali2013} and effective fermion masses $m_f$~\cite{Dvali2016a} emerge at a new fundamental energy scale. Note that this is true for arbitrary numbers of massless fermion flavors, as it is enforced by gravitational anomaly matching conditions~\cite{Dvali2017}. Thus, we get
\begin{equation}
\Lambda_G = \langle R\tilde{R},R\tilde{R}\rangle_{q\to 0}^{1/8}\sim |\langle \bar{f}f\rangle|^{1/3}\sim m_f.
\end{equation}
Phenomenological constraints push this scale into the range $\Lambda_G\sim 0.1$~meV--eV~\cite{Dvali2016a, Dvali2016b,Funcke:2019grs} (see Ref.~\footnote{The scale $\Lambda_G$, the neutrino condensate $\langle\bar{\nu_i}\nu_i\rangle$, and the temperature $T_{\Lambda_G}$ of the phase transition are free parameters of the theory, which can differ by unknown $\mathcal{O}(10^{-1}-10)$ parameters and are fixed by phenomenological constraints, including: (i) corrections to Newtonian gravity are experimentally excluded up to $\langle\bar{\nu_i}\nu_i\rangle\gtrsim {\rm meV}$~\cite{Kapner2006}; (ii) neutrino mass splittings require the largest neutrino condensate to be $\langle\bar{\nu_i}\nu_i\rangle\geq 4$~meV due to $m_{\nu_i}=g_i\langle\bar{\nu_i}\nu_i\rangle$ and $g_i\leq 4\pi$~\cite{ParticleDataGroup:2020ssz};
(iii) neutrino free-streaming before photon decoupling requires $T_{\Lambda_G}\lesssim T_{\rm CMB}=0.3$~eV~\cite{Dvali2016a}. Note that the constraints (i) and (iii) are model-independent constraints, which still apply if neutrinos get small masses through other mechanisms. Due to the unknown $\mathcal{O}(10^{-1}-10)$ parameters, the scale $\Lambda_G$ can take values between $\mathcal{O}(0.1~{\rm meV})$ and $\mathcal{O}({\rm eV})$~\cite{Dvali2016a}.} for more details). This opens up the possibility that neutrino masses $m_{\nu_i}$ are generated through a Higgs-like composite field $\langle\Phi\rangle\equiv\langle \bar{\nu_i}\nu_i\rangle$ at low energies~\cite{Dvali2016a} instead of high-energy extensions of the SM. 

In this Letter, we will study the topological defects that form in the late Universe due to this neutrino condensation. The model~\cite{Dvali2016a} predicts a phase transition in the late Universe after recombination, in which neutrino flavor symmetry gets spontaneously broken, small effective neutrino masses $m_\nu\sim\Lambda_G$ arise, and pseudoscalar Nambu-Goldstone bosons $\phi$ appear~\cite{Dvali2013,Dvali2016a,Dvali2016b}. Moreover, topological defects form due to the Kibble mechanism~\cite{Kibble1976}, including skyrmions, global monopoles, strings, and domain walls (DWs). These defects are very soft, only require SM particle content, and depend on the Majorana or Dirac nature of the neutrinos. In the current work, we will provide a detailed study of the formation, evolution, and observational consequences of these defects. Beyond these observational consequences presented here, two more experimental signatures of this model are post-recombination relic neutrino annihilation and astrophysical neutrino decay~\cite{Dvali2016a, Dvali2016b,Funcke:2019grs}. The former prediction can be experimentally verified or falsified with the next-generation cosmological surveys, while the latter can be tested with future astrophysical neutrino observatories.

\textit{Relic neutrino phase transition.}---The low-energy neutrino mass model proposed in Ref.~\cite{Dvali2016a} predicts that the cosmological neutrinos remain massless until the very late Universe, $T\lesssim T_{\rm CMB}\sim 0.3$~eV. At that time, a neutrino vacuum condensate forms in a cosmological phase transition triggered by the gravitational vacuum susceptibility. The condensation generates small effective neutrino masses and (pseudo)Goldstone bosons, $\phi\equiv\{\phi_k,\eta_\nu\}$, which are called ``gravi-majorons'' in Ref.~\cite{Funcke:2019grs}. 
These neutrino-composite bosons arise from the spontaneous breaking of the neutrino flavor symmetry, which reads $U(3)_L$ in the case of left-handed Majorana neutrinos (for which $k=1,...,8$) and $U(3)_L\times U(3)_R$ in the case of Dirac neutrinos (for which $k=1,...,14$); see Eqs.~\eqref{eq:MInitialSymm} and~\eqref{eq:DInitialSymm}. 
 
The $\eta_\nu$ boson is analogous to the heavy $\eta'$ meson of QCD and acquires a small mass through the chiral gravitational anomaly, $m_{\eta_\nu}\sim \Lambda_G$~ \citep{Dvali2013}. At first sight, one might expect the $\phi_k$ bosons to be massless, since the neutrinos in the model \citep{Dvali2016a} have no hard masses but only effective ones. However, loop diagrams involving $W$-boson and charged-lepton exchange give rise to tiny mass contributions for some of the $\phi_k$ bosons [see Eq.~(12) in the Supplemental Material~\cite{sup_mat}],
\begin{equation}\label{eq:mass}
m_{\phi_k}=\frac{G_F}{4\pi} m_l\, \Lambda_G^2=2\times 10^{-21}~{\rm eV}\left(\frac{m_l}{m_\tau}\right)\left(\frac{\Lambda_G}{\rm 1~meV}\right)^2,
\end{equation}
thus explicitly breaking the original $U(3)_L$ neutrino flavor symmetry. The largest (smallest) contributions to the $\phi_k$-boson masses are given for $m_l=m_\tau$ ($m_l=m_\mu$). 

After the phase transition, the massive relic neutrinos rapidly decay into the lightest neutrino mass eigenstate, $\nu_i\rightarrow\nu_j+\phi_k$, 
and bind up or annihilate into the (almost) massless $\phi_k$ bosons, $\nu_j + \bar{\nu}_j\rightarrow \phi_k + \phi_k$~\cite{Dvali2016a}. Note that these bosons only form in the very late Universe and thus do not alter any early-Universe physics.
Almost all neutrinos convert into this dark radiation in the late Universe, apart from a negligibly small freeze-out density. This late ``neutrinoless Universe'' scenario can only 
be evaded in the hypothetical presence of neutrino asymmetries~\cite{Lorenz:2018fzb}.
The absolute neutrino mass scale is constrained by $m_\nu\lesssim T_{\rm CMB}\sim 0.3$ eV if the phase transition takes place instantaneously at a temperature $T_{\Lambda_G}\sim \Lambda_G \sim m_\nu$~\cite{Dvali2016a,Koksbang2017}
so as not to conflict with cosmic microwave background (CMB) observations. However, the phase transition can also 
be supercooled and thus can give rise to relatively large neutrino masses even at a low apparent transition temperature, 
$T_{\Lambda_G}< \Lambda_G\sim m_\nu \lesssim 1.5$~eV at 95\%~C.L.~\cite{Lorenz:2018fzb,Lorenz:2021alz}. Such a supercooling mechanism would allow for 
substantial energy densities of the $\phi_k$ bosons and the topological defects after the 
transition.

In the following, we will discuss the neutrino flavor symmetry breaking patterns from neutrino condensation. We will first focus on the case that neutrinos are Majorana particles, followed by a discussion of the Dirac case. We will restrict our analysis to the minimal SM scenario of three active neutrino species. However, we wish to emphasize that there could be additional sterile neutrino species, which could enhance the neutrino flavor symmetry. Because of the universality of the gravitational neutrino mass mechanism~\cite{Dvali2016a}, the topological structure in the neutrino sector would be sensitive to the structure of such hidden-sector sterile neutrinos. Interestingly, because of the unique role of neutrinos as the lightest and weakest interacting particles of the SM, this offers an exciting new opportunity to probe hidden-sector physics.

\textit{Symmetry breaking in the Majorana case.}---The minimal version of the SM only contains three left-handed neutrinos $\nu_{L\alpha}$ (where $\alpha=e,\mu,\tau$ is the flavor index) and no right-handed neutrinos. Thus, the minimal version of the gravitational neutrino mass model~\cite{Dvali2016a} generates effective Majorana masses for these three left-handed species.

When neglecting SM interactions, the Lagrangian of the three massless neutrinos only contains a kinetic term,
\begin{equation}
\mathcal{L}_\nu= i{\bar \nu}_{L\alpha} \slashed{\partial} \nu_{L\alpha} + {\rm h.c.},
\label{eq:LnuMaj}
\end{equation}
where the implicit sum over the flavor index $\alpha = e, \mu, \tau$ could be equivalently written as a sum over the mass index $i = 1,2,3$ [just as in Eq.~\eqref{eq:LnuDir} below]. The Lagrangian in Eq.~\eqref{eq:LnuMaj} has the flavor symmetry
\begin{align}
\begin{split}
\label{eq:MInitialSymm}
G &= U(3)_L= \frac{SU(3)_L\times U(1)_L}{Z_{3}}.
\end{split}
\end{align}
This neutrino flavor symmetry is a quotient group because the $SU(3)$ and $U(1)$ symmetries have a common center, $U(3)=[SU(3)\times U(1)]/Z_{3}$.
On top of the small explicit breaking through weak effects [Eq.~\eqref{eq:mass}], the flavor symmetry in Eq.~\eqref{eq:MInitialSymm} gets explicitly broken by the chiral gravitational anomaly~\cite{Delbourgo1972,Eguchi1976, Deser1980,AlvarezGaume1984} and spontaneously by the neutrino condensate~\cite{Dvali2016a}. The explicit breaking of the $U(1)_L$ part in Eq.~\eqref{eq:MInitialSymm} is similar to invisible axion scenarios~\cite{Peccei1977,Weinberg1977,Wilczek1978,Kim1979, Shifman1979, Zhitnitsky1980,Dine1981}, where the $U(1)_{\rm PQ}$ symmetry is explicitly broken by the chiral 
anomalies of gravity and QCD. For the spontaneous breaking, 
the order parameter reads
\begin{equation}
\Phi_{\alpha\beta}\equiv \nu_{L\alpha}^T C\nu_{L\beta},
\end{equation}
where $C=i\gamma^2\gamma^0$ is the charge conjugation matrix. Note that the order parameter is symmetric, 
$\Phi_{\alpha\beta}=\Phi_{\beta\alpha}$. 

The effective potential of the symmetry breaking reads
\begin{align}
\begin{split}
V(\Phi)=&\,\,a \Tr(\Phi^\dagger\Phi)+b\Tr(\Phi^\dagger\Phi\Phi^\dagger\Phi)+\cdots\\
&+c \det \Phi+ {\rm h.c.}+\cdots,
\end{split}
\label{eq:potential}
\end{align}
where $a$, $b$, and $c$ are unknown numerical coefficients and $\det \Phi$ is a 't Hooft-like determinant~\citep{'tHooft1986} which breaks the $U(1)_L$ symmetry in Eq.~\eqref{eq:MInitialSymm} down to $Z_6$. Note that the potential in Eq.~\eqref{eq:potential} does not necessarily need to respect the full $U(3)$ symmetry but could minimally be invariant under $O(3)$ symmetry only, because the maximal flavor symmetry of flavor-invariant Majorana masses is $O(3)$.

Gravitational anomaly matching conditions require all neutrinos to become massive~\cite{Dvali2017}, and these masses need to be different due to the observed neutrino flavor oscillations. Thus, in the neutrino mass eigenbasis, the vacuum expectation value reads
\begin{equation}
\langle \Phi\rangle = \begin{pmatrix}
\langle\Phi_{11}\rangle & 0 & 0\\
0 & \langle\Phi_{22}\rangle & 0\\
0 & 0 & \langle\Phi_{33}\rangle
\end{pmatrix},
\label{eq:MVEV}
\end{equation}
where $\langle\Phi_{11}\rangle\neq\langle\Phi_{22}\rangle\neq\langle\Phi_{33}\rangle$ for different masses. 

In order to determine the resulting flavor symmetry breaking pattern, we need to find all elements of the initial group $G$ in Eq.~\eqref{eq:MInitialSymm} that leave $\langle\Phi_{11}\rangle$, $\langle\Phi_{22}\rangle$, and $\langle\Phi_{33}\rangle$ invariant, which gives us the unbroken symmetry group $H$. 
Among the elements of $SU(3)_L\times U(1)_L$, only the $Z_2$ elements $\{\mathbb{1},e^{i\pi\mathbb{1}}\}$ leave the different condensates $\langle\Phi_{ii}\rangle$ invariant (for each $i=1,2,3$, with no sum over repeated indices). 
Therefore, we finally get the symmetry group $H = Z_2\times Z_2\times Z_2$ from the spontaneous breaking
\begin{equation}
SU(3)_L\rightarrow Z_2\times Z_2,\quad U(1)_L\rightarrow Z_2, 
\label{eq:U(1)L-Z2}
\end{equation}
where the elements of $Z_2\times Z_2$ are given by
\begin{align}
\begin{split}
&g_1 ={\rm diag}(1,-1,-1),\quad g_2 ={\rm diag}(-1,1,-1),\\
&g_3 ={\rm diag}(-1,-1,1),\quad g_4 =\mathbb{1}.
\end{split}
\label{eq:z2z2}
\end{align}
Note that this $Z_2\times Z_2$ symmetry is explicitly broken by weak effects at very low energy scales; see Eq.~\eqref{eq:mass}.

\textit{Symmetry breaking in the Dirac case.}---As a minimal extension of the SM, the three left-handed neutrinos $\nu_{L\alpha}$ could be accompanied by two or more right-handed partners. In the following, we will focus on the three-flavor Dirac neutrino case of the neutrino mass model~\cite{Dvali2016a}, noting that the model also allows for mixed active-sterile neutrinos with both Dirac and Majorana masses. 

When neglecting SM interactions, the Lagrangian of the three massless neutrinos only contains a kinetic term,
\begin{equation}
\mathcal{L}_\nu=i{\bar \nu}_{L\alpha} \slashed{\partial} \nu_{L\alpha} + i{\bar \nu}_{R\alpha} 
\slashed{\partial} \nu_{R\alpha} = i{\bar \nu}_{D\alpha} \slashed{\partial} \nu_{D\alpha} .
\label{eq:LnuDir}
\end{equation}
In the limit in which all interactions except gravity are switched off, the Lagrangian has a $U(6)$ flavor symmetry. When taking into account the small explicit breaking by the weak interaction [Eq.~\eqref{eq:mass}], 
the left- and right-handed sectors cannot mix, and we can write the symmetry as
\begin{align}
\begin{split}
\label{eq:DInitialSymm}
G &= U(3)_L\times U(3)_R= \frac{U(3)_V\times U(3)_A}{Z_2} \\
&= \frac{SU(3)_V\times SU(3)_A\times U(1)_V\times U(1)_A}{Z_{V3}\times Z_{A3}\times Z_2},
\end{split}
\end{align}
where $L$, $R$, $V=L+R$, and $A=L-R$ denote the left, right, vector, and axial symmetries, respectively. The $L$ and $R$ group elements act on $\nu_D$ and take the form $\exp[i(1\pm \gamma_5) \alpha_a T_a]$, 
where the $T_a$ are group generators.
The flavor symmetry $G$ in Eq.~\eqref{eq:DInitialSymm} is a quotient group because (i) the $SU(3)$ and $U(1)$ symmetries have a common $Z_3$ center and (ii) the $\pi$ rotations of the $U(1)_V$ and $U(1)_A$ symmetries are the same element, $\exp(i\pi)=\exp(i\pi\gamma_5)=-1$, which yields $U(1)_L\times U(1)_R=[U(1)_V\times U(1)_A]/Z_2$.

Similar to the Majorana case [Eq.~\eqref{eq:MInitialSymm}], the initial flavor symmetry $G$ in Eq.~\eqref{eq:DInitialSymm} is explicitly broken by the gravitational anomaly and spontaneously by the neutrino condensate~\cite{Dvali2016a}. 
The order parameter in the Dirac case is
\begin{equation}
\Phi^D_{ij}\equiv {\bar \nu}_i\nu_j
\end{equation}
in the mass eigenbasis, 
and its expectation value
$\langle\bar{\nu}_i\nu_j\rangle$
is bifundamental and obeys $\langle\bar{\nu}_i\nu_i\rangle\neq \langle\bar{\nu}_j\nu_j\rangle$ for $i\ne j$. 
The effective potential is similar to the one in Eq.~\eqref{eq:potential}.

Among the elements of $SU(3)_A\times U(1)_A$, only the $Z_2$ elements 
$\{\mathbb{1},e^{i\pi\gamma_5\mathbb{1}}\}$ leave $\langle\bar{\nu}_i\nu_i\rangle$
invariant (for each $i=1,2,3$). 
Therefore, we get
\begin{equation}
SU(3)_A\rightarrow \mathbb{1},\quad U(1)_A\rightarrow Z_2.
\label{eq:U(1)A-Z2}
\end{equation}
Here, we neglected the explicit symmetry breaking 
of $U(1)_{A}\rightarrow Z_6$ [just as we neglected the explicit breaking of $U(1)_{L}$ in Eq.~\eqref{eq:U(1)L-Z2}] due to the 't Hooft-like determinant~\eqref{eq:potential}, 
as discussed in detail in Ref.~\cite{sup_mat}.

Regarding the $U(3)_V$ symmetry, any phase transformation of $\nu_i$ leaves $\langle\bar{\nu}_i\nu_i\rangle$ invariant. Therefore, we get three unbroken $U(1)$ symmetries, 
\begin{equation}
U(3)_V\rightarrow  U(1)_{V3}\times \frac{U(1)_{V1}\times U(1)_{V8}}{Z_{V3}},
\label{eq:U^3}
\end{equation}
where the elements of $U(1)_{V1}$ are $e^{i\alpha \mathbb{1}}$, of $U(1)_{V3}$ are $e^{i\alpha\lambda_3}$, and of $U(1)_{V8}$ are $e^{i\alpha\lambda_8}$, with $\lambda_3={\rm diag}(1,-1,0)$
and $\lambda_8={\rm diag}(1,1,-2)/\sqrt{3}$. 
 The $Z_{V3}$ denominator is because the center of $SU(3)_V$ is given by $\{\mathbb{1}, e^{i2\pi/3}\mathbb{1},e^{i4\pi/3}\mathbb{1}\}$ and is contained in both $U(1)_{V1}$ and $U(1)_{V8}$.
Thus, the symmetry breaking is effectively
\begin{equation}
SU(3)_V\rightarrow U(1)_{V3}\times U(1)_{V8}
\label{eq:U^2}
\end{equation}
and the unbroken symmetry group reads
$H=U(1)_{V3}\times[U(1)_{V1}\times U(1)_{V8}]/Z_{V3}$ because the $Z_2$ denominator in Eq.~\eqref{eq:DInitialSymm} cancels the numerator in Eq.~\eqref{eq:U(1)A-Z2}. 

From a broader perspective, we observe that the symmetry breaking patterns substantially differ for the Majorana and Dirac cases. Thus, the resulting topological defects crucially depend on the yet unknown neutrino nature and offer the exciting possibility of shedding new light onto this nature. In the following, we show that only the Majorana neutrino case yields topological defects that can induce a time- and space-varying neutrino mass matrix. In the Dirac case, the defects are quickly annihilating and thus are not expected to have any observational consequences, as demonstrated  
in Ref.~\cite{sup_mat}.

\textit{Cosmic string network.}---In the Majorana case, the original $SU(3)_L$ flavor symmetry of the massless left-handed Majorana neutrinos is spontaneously broken down to $Z_2\times Z_2$; see Eq.~\eqref{eq:U(1)L-Z2}. This symmetry breaking gives rise to global strings, 
\begin{equation}
\pi_1[SU(3)/(Z_2\times Z_2)]=Z_2\times Z_2,
\label{eq:Z2xZ2}
\end{equation}
which are analogous to the cosmic strings of broken flavor symmetry first discussed in Refs.~\cite{Bibilashvili:1990qm, Dvali:1991ka, Dvali:1993qp}. The order parameters in the current and original flavor string scenarios are similar, except that in the present case it is composite instead of fundamental.

When investigating the cosmological evolution of the string network,
we obtain the characteristic length scale $\xi$ of the strings [see Eq.~(18) in Ref.~\cite{sup_mat}], which reads
\begin{align}
\begin{split}
\xi =\sqrt{\frac{\lambda^2\Lambda_G^7 t}{a_G^2 T_\nu^8}}= 10^{14}~{\rm m} \left(\frac{\lambda}{1}\right)\left(\frac{\Lambda_G}{1~{\rm meV}}\right)^{\frac{7}{2}}\left(\frac{1}{a_G}\right).
\end{split}
\label{eq:finalxi}
\end{align}
Here, $\lambda$ is the self-coupling of the scalar string field~$\Phi$, $\Lambda_G$ is the infrared gravitational scale, $t$ is the Hubble time, $a_G$ is the scale factor of the phase transition, and $T_\nu\sim 0.2~{\rm meV}$ is the neutrino temperature. 
This scale sets the distance over which the strings are straight, which 
is also assumed to be the inter-string separation. 

When traveling around a string, the vacuum expectation value $\langle\Phi\rangle$ winds continuously by the $SU(3)$
group transformation
\begin{align}
\begin{split}
\langle\Phi(\theta)\rangle = \omega^T(\theta)
\langle\Phi\rangle
\omega(\theta),
\end{split}
\label{Phirotation}
\end{align}
where $\langle\Phi\rangle$ is defined in Eq.~\eqref{eq:MVEV}. While the angle $\theta$ runs from $0$ to $2\pi$, the path $\omega(\theta)$ interpolates between different elements $g_i\in Z_2\times Z_2$ [Eq.~\eqref{eq:z2z2}], which label the different strings~\cite{Mermin:1979zz}. For example, the string labeled by the $g_3$ element 
corresponds to a path $\omega(\theta)$ that has the form
\begin{equation}
\omega_3(\theta)
=\begin{pmatrix}
\cos\left({\theta}/{2}\right) & \sin\left({\theta}/{2}\right) & 0\\
-\sin\left({\theta}/{2}\right) & \cos\left({\theta}/{2}\right) & 0\\
0 & 0 & 1
\end{pmatrix},
\label{eq:omega3}
\end{equation}
which corresponds to a rotation around a third axis in flavor space. Thus, when a neutrino passes by a string, the mixing angles of the standard Pontecorvo-Maki-Nakagawa-Sakata (PMNS) neutrino mixing matrix~\cite{Pontecorvo:1957qd,Maki:1962mu} change nontrivially. The PMNS matrix $U_{\rm PMNS}=U^\dagger_e U_\nu$ is a product of two unitary matrices $U_e$ and $U_\nu$ arising from the diagonalization of the charged lepton and neutrino mass matrices. Keeping $U_e$ fixed, the matrix $U_\nu$ changes by $\omega(\theta)U_\nu$ when passing by a string, resulting in $U_{\rm PMNS}=U^\dagger_e\omega(\theta) U_\nu$. 
Thus, the parameters of the PMNS matrix and the resulting strength of $CP$ violation (given by a product of these parameters) are time dependent and space dependent.
As discussed in Refs.~\cite{sup_mat,Bibilashvili:1990qm, Dvali:1991ka, Dvali:1993qp}, the strings can only induce such observable changes of the neutrino mass matrix if the residual $Z_2\times Z_2$ symmetry in Eq.~\eqref{eq:Z2xZ2} is further broken. This breaking gives rise to DWs, as we will demonstrate in the following section.

\textit{Topological domain walls.}---In the original flavor string models 
in Refs.~\cite{Bibilashvili:1990qm, Dvali:1991ka, Dvali:1993qp}, the residual $Z_2\times Z_2$ symmetry 
in Eq.~\eqref{eq:Z2xZ2} needed to be further broken 
at high-energy scales, $\Lambda_{Z_2}\sim (10$--$100)$~MeV, in order to ensure sizable lepton mass 
differences, $m_e\neq m_\mu$. In our case, the $Z_2\times Z_2$ symmetry in Eq.~\eqref{eq:U(1)L-Z2} is explicitly broken at the low-energy scales given by Eq.~\eqref{eq:mass}, such that the resulting 
topological DWs have a width of
\begin{equation}
\delta_{\rm DW}=\frac{1}{m_{\phi_k}}= 8\times 10^{14}~{\rm m}\left(\frac{m_\tau}{m_l} \right)\left(\frac{1~{\rm meV}}{\Lambda_G} \right)^2,
\label{eq:DWwidth}
\end{equation}
where $m_l$ is either $m_\tau$ or $m_\mu$, depending on the $\phi_k$ boson under consideration. Thus, there are two independent string-wall networks, in which the DWs of width $\delta_{\rm DW}(m_l=m_\mu)$ and $\delta_{\rm DW}(m_l=m_\tau)$ are connected to the two different types of $Z_2$ strings in Eq.~\eqref{eq:Z2xZ2}, respectively. The energy density in these DWs is $\rho_{\rm DW} = \delta_{\rm DW}^3\xi$.

As discussed above, the explicit $Z_2\times Z_2$ symmetry breaking that gives rise to the DWs is the reason why passing by a string results in observable neutrino flavor transitions. As an example, we can consider a DW 
bounded by a $g_3$ string, described by $\omega_3(\theta)$ in Eq.~\eqref{eq:omega3}.  As we go around the string, $\langle \Phi (\theta)\rangle$ is rotated by $\omega_3^T(\theta)
\langle\Phi\rangle
\omega_3(\theta)$ as
in Eq.~\eqref{Phirotation} and returns to itself for $\theta=2\pi$. Meanwhile, the neutrino wave function rotates by $\omega_3(\theta)\xrightarrow[\theta=2\pi]{\text{}}{\rm diag}(g_3,1)$, which leads to an overall minus sign of $\nu_1$ and $\nu_2$.
Such a DW can arise through the vacuum expectation value (VEV) $\langle \Phi \rangle \propto (1,0,0)$, for which 
$g_3  \langle \Phi\rangle\neq \langle \Phi \rangle$ 
with $g_3$ given by Eq.~\eqref{eq:z2z2}. 
Thus, the neutrinos 
change their flavor when passing by a string or through a DW.

\textit{Experimental predictions.}---In principle, both the DWs and the strings that form in the Majorana neutrino case can induce time- and space-dependent effects on the neutrino mass matrix. However, the DW effects are negligibly small, as they only change the neutrino mass splittings by $\mathcal{O}(\delta_{\rm DW}^{-1})\lesssim 2\times 10^{-21}$~eV; see Eq.~\eqref{eq:mass}. While the string effects are larger, they do not change the neutrino mass splittings but only the mixing angles, as explained below Eq.~\eqref{eq:omega3}. While a neutrino travels from one string to another, the mixing angles continuously and randomly change by $\mathcal{O}(1)$, which is the leading observational effect of the topological defects. As a secondary effect, the strings induce different mixing angles at Sun and Earth, but this effect is negligibly small because the Earth-Sun distance is smaller than the inter-string separation.

The time-dependent change of the neutrino mixing angles can be observable in future neutrino experiments and already in existing neutrino data. Similar experimental signatures have been previously discussed in the context of dark-matter-neutrino interactions, which can give rise to time-dependent oscillatory contributions to the neutrino masses and mixing angles~\cite{Berlin:2016woy, Krnjaic:2017zlz, Brdar:2017kbt, Dev:2020kgz, Losada:2021bxx}. However, the predictions of our scenario fundamentally differ from the previously considered ones, because only the mixing angles are time dependent. Interestingly, current data allow for an $\mathcal{O}(10\%)$ variation in the neutrino parameters over the last ten years~\cite{ParticleDataGroup:2010dbb,ParticleDataGroup:2020ssz}.

To determine the timescales over which the neutrino mixing angles change, we need to compare the inter-string separation scale $\xi$ in Eq.~\eqref{eq:finalxi} to the DW width $\delta_{\rm DW}$ in Eq.~\eqref{eq:DWwidth}. 
Here, we can distinguish two scenarios: 
For $\xi>\delta_{\rm DW}$, the DWs would dominate the evolution of the string-wall network, and the network would dilute rapidly, similar to the axionic and non-topological DWs treated in Ref.~\cite{sup_mat}, with $\mathcal{O}(10)$ DWs remaining per Hubble volume. In this case, an observation of variations in the mixing angles would be very unlikely. We thus focus only on the case $\xi<\delta_{\rm DW}$ in the following.

Next, we need to compare the distance scale $\xi$ and the timescale $t=\xi/v$ to the corresponding scales of the neutrino experiments under consideration. For example, experiments like Daya Bay~\cite{DayaBay:2018yms}
measured $\sin^2(2\theta_{13}) =  0.0856 \pm 0.0029$ taking data for $t\sim 6$~years. During this time, the solar system moved a distance $d=vt= 4\times 10^{13}$~m through the frozen string and DW background, where $v\sim 230$~km/s. Interestingly, this distance is similar to the inter-string separation $\xi$ in Eq.~\eqref{eq:finalxi} and therefore similar to the distance over which the neutrino mixing angles change by $\mathcal{O}(1)$. This implies that current data already exclude the existence of the smallest possible inter-string separation $\xi<\delta_{\rm DW}=8\times 10^8$~m for $m_l=m_\tau$ and $\Lambda_G=1$~eV. In this case, the Solar System would only need $t=d/v\sim 1$~h to pass by a string,
which would result in rapid variations of the neutrino mixing angles.
While this specific case of $\xi<\delta_{\rm DW}$ and $\Lambda_G=1$~eV is ruled out, both the cases of $\xi>\delta_{\rm DW}$ and $\xi<\delta_{\rm DW}$ with $\Lambda_G\lesssim 0.2$~meV are still viable. In the latter case, the most constraining parameter is the smallest mixing angle $\sin^2(2\theta_{13})=0.0856 \pm 0.0029$~\cite{DayaBay:2018yms}, which could exhibit a time-dependent change $\Delta_{13}$ within the experimental uncertainty, $\Delta_{13}=\sin^2(2\times 2\pi vt/\delta_{\rm DW})\lesssim 2\times 0.0029$, where $t\sim 6$~y and $\delta_{\rm DW}$ is given by Eq.~\eqref{eq:DWwidth}. This estimate strongly motivates the search for time-dependent changes in existing and future neutrino oscillation data, with a particular focus on the smallest mixing angle $\theta_{13}$.

\textit{Conclusions.}---We studied the formation and evolution of soft topological defects from a late cosmological phase transition predicted by the neutrino mass model in Ref.~\cite{Dvali2016a}. In the model, neutrino flavor symmetry gets spontaneously broken by a composite Higgs-like field, which is neutrino bilinear and thus does not require any new particles. 
We demonstrated that the Dirac case gives rise to neutrino skyrmions, global monopoles, and a hybrid string-wall network, which all quickly annihilate into dark radiation. More importantly, the Majorana case predicts the formation of strings and topological DWs that are similar to the strings and DWs of broken flavor symmetry first discussed in Refs.~\cite{Bibilashvili:1990qm, Dvali:1991ka, Dvali:1993qp}. This string-wall network induces a time- and space-dependent variation of the neutrino mixing angles, which could be observable in future neutrino experiments and already in existing neutrino data. The detection of this smoking gun feature would indicate that neutrinos are Majorana particles and thus could solve one of the fundamental open questions of neutrino physics.\\

\begin{acknowledgments}

We thank Cecilia Lunardini, Georg Raffelt, Goran Senjanovi\'{c}, Alexei Smirnov, and Alex Vilenkin for insightful discussions and the referees for valuable remarks. G.D.\ is supported in part by the Humboldt Foundation under Humboldt Professorship Award, by the Deutsche Forschungsgemeinschaft (DFG, German Research Foundation) under Germany’s Excellence Strategy - EXC-2111 - 390814868, and Germany’s Excellence Strategy under Excellence Cluster Origins. L.F.\ is supported by the DOE QuantiSED Consortium under Subcontract No.\ 675352, by the National Science Foundation under Cooperative Agreement PHY-2019786 (The NSF AI Institute for Artificial Intelligence and Fundamental Interactions~\citep{IAIFI}), and by the U.S.\ Department of Energy, Office of Science, Office of Nuclear Physics under Grant Contracts No.\ DE-SC0011090 and No.\ DE-SC0021006. T.V.\ is supported by the U.S.\ Department of Energy, Office of High Energy Physics, under Award No.\ DE-SC0019470 at ASU.
\end{acknowledgments}

\appendix
\onecolumngrid
\vspace{\columnsep}
\begin{center}
\textbf{\large \textbf{\large Supplemental Material: Details on Goldstone Boson Masses}\\*[0.2em] and Topological Defects Without Observational Consequences}
\end{center}
\vspace{0.6\columnsep}
\twocolumngrid

\section{Goldstone boson masses}\label{app:goldstones}

In this section, we compute the $\phi_k$-boson masses that are generated through loop diagrams involving $W$-boson and charged-lepton exchange. 
These masses determine the topological domain wall widths in Eq.~(22) of the main text and have a crucial impact on the observable consequences of the cosmic strings. To demonstrate the latter, let us in the following first discuss a toy model in which these $\phi_k$-boson masses are absent. 

For this discussion, we use a slightly different notation as in the main paper. In particular, we do not use $\alpha$ as a neutrino flavor index and $i$ as a neutrino mass index, but we assume that the mass matrices of the leptons ($\nu_i$ and $e_\alpha$, see below) are aligned at $\theta=0$. Thus, the flavor and mass indices are equivalent and there is no mixing in the lepton sector at $\theta=0$. Below Eq.~\eqref{eq:mnutheta}, we then discuss how mixing arises for nonzero values of $\theta$. 

In the toy model, we have two Majorana neutrinos $\nu_i$ with flavor (or mass) index $i=1,2$. These neutrinos gain masses through the condensation of a scalar field $\phi_{ij}$ that is in a symmetric, reducible representation of the $SO(2)$ neutrino flavor symmetry. For simplicity, we assume the field $\phi_{ij}$ to be fundamental instead of composite, which is different from the model discussed in the main paper. Further, the toy model contains a gauge boson $W_\mu$ similar to the one in the Standard Model and two additional fermions $e_\alpha$ with flavor (or mass) index $\alpha=1,2$. The latter are similar to the electron and muon of the Standard Model, but are assumed to be electrically neutral Majorana fermions instead of Dirac fermions. Moreover, as mentioned above, we first neglect lepton mixing for simplicity and assume that the mass matrices of $\nu_i$ and $e_\alpha$ are diagonal in the same basis.

The Lagrangian of this toy model reads
\begin{align}
\begin{split}
\mathcal{L}_{\rm toy} &= \nu_i^T\slashed{\partial}\nu_i + e_\alpha^T\slashed{\partial}e_\alpha-\frac{1}{4}F_{\mu\nu}F^{\mu\nu}+m_W^2 W_\mu W^\mu \\
&+ m_\alpha^e e_\alpha^T\gamma^0 e_\alpha +{\rm Tr} \partial_\mu \phi_{ij}\partial^\mu\phi_{ij} - \lambda^2({\rm Tr}\phi_{ij}^2-v^2)^2 \\
&+ g\phi_{ij}\nu_i^T\gamma^0\nu_j + g_W W_\mu l_\beta^T\gamma^0\gamma^\mu l_\beta,
\end{split}
\label{eq:toyLag}
\end{align}
where $F_{\mu\nu}$ is the field strength tensor and $m_W$ is the mass of the $W_\mu$ field, $l_\beta$ is the weak lepton doublet ($\beta=1,2$), $m_\alpha^e$ are the electron and muon masses, $g$ is the neutrino-scalar coupling, $g_W$ is the weak coupling, and $v$ is the vacuum expectation value of the $\phi_{ij}$ field that spontaneously breaks the $SO(2)$ neutrino flavor symmetry and generates the neutrino masses.

In the limit of $m_\alpha^e\to 0$, there is a full $SO(2)$ symmetry that is spontaneously broken, and both $(\nu_1, \nu_2)^T$ and $(e_1, e_2)^T$ transform as doublets under $SO(2)$. In the limit $g_W\to 0$, only $(\nu_1, \nu_2)^T$ transforms under $SO(2)$. In the case $g_W\neq 0$ and $m_1^e\neq m_2^e$, there is an approximate $SO(2)$ symmetry for $(\nu_1, \nu_2)^T$, which is broken by the mass difference $|m_1^e- m_2^e|\neq 0$.

Now, let us consider the case $g_W= 0$ and $m_1^e\neq m_2^e$ and study the cosmic strings that form in the transition,
\begin{align}
\begin{split}
\langle\phi\rangle &= \omega_3^T(\theta)v\,\omega_3(\theta) \\
&= v
\begin{pmatrix}
\cos^2\left({\theta}/{2}\right) & \cos\left({\theta}/{2}\right)\sin\left({\theta}/{2}\right) \\
\cos\left({\theta}/{2}\right)\sin\left({\theta}/{2}\right) & \sin^2\left({\theta}/{2}\right)
\end{pmatrix},
\end{split}
\label{eq:toystring}
\end{align} 
where $\omega_3(\theta)$ is the upper left $2\times2$ matrix of the analogous three-flavor matrix given in Eq.~(21) of the main text. 

If $g_W= 0$, the strings do not have any observable consequences. In order to see this, let us insert Eq.~\eqref{eq:toystring} into the neutrino-dependent part of Eq.~\eqref{eq:toyLag},
\begin{align}
\begin{split}
\mathcal{L}_{\rm toy, \nu} &= \nu_i^T\slashed{\partial}\nu_i + g\phi_{ij}\nu_i^T\gamma^0\nu_i \\
&= \nu_i^T\slashed{\partial}\nu_i + gv(\nu_1^T,\nu_2^T)
\begin{pmatrix}
c^2 & cs \\
cs& s^2
\end{pmatrix}\gamma^0
\begin{pmatrix}
\nu_1 \\
\nu_2 
\end{pmatrix},
\end{split}
\label{eq:toynu}
\end{align}
where $c^2\equiv \cos^2(\theta/2)$, etc. When a neutrino goes around the string, the first neutrino species $\nu_1$ has a nonzero mass $m_1=vg$ at $\theta=0$ but a zero mass at $\theta=\pi$, while the opposite is true for the second neutrino species $\nu_2$,
\begin{align}
m_{\nu_i}(\theta=0)= gv
\begin{pmatrix}
1 & 0 \\
0& 0
\end{pmatrix}
\,\,\to\,\,
m_{\nu_i}(\theta=\pi)= gv
\begin{pmatrix}
0 & 0 \\
0& 1
\end{pmatrix}.
\label{eq:mnutheta}
\end{align}
As mentioned before, we assume that the mass matrices of $\nu_i$ and $e_\alpha$ are diagonal in the same basis at $\theta=0$, such that the neutrino flavor and neutrino mass indices are equivalent. We also note that without coupling the neutrinos to the electrons and muons via $g_W\neq 0$, the definition of the neutrino species $\nu_1$ and $\nu_2$ in Eq.~\eqref{eq:toynu} is arbitrary. Thus, without a reference point, the statement that we produce the massive $\nu_1$ state at $\theta=0$ and measure the massless state $\nu_2$ at $\theta=\pi$ is empty. Only when turning on $g_W\neq 0$, we can define $\nu_1$ as the neutrino that is coupled to $e_1W_\nu$, such that changing the massive neutrino from $\nu_1$ at $\theta=0$ to $\nu_2$ at $\theta=\pi$ in Eq.~\eqref{eq:mnutheta} by going around the string becomes an observable effect.

The observability of changing the neutrino mass matrix while going around the string is directly connected to the nonzero masses of the $\phi_k$ bosons, as we will demonstrate in the following. For $g_W\neq 0$, the $\phi_k$ bosons gain masses through loop diagrams, which differ depending on the Majorana or Dirac nature of the electron and muon species. In our toy model, the electron and muon are Majorana fermions, which allows for two-loop diagrams with two external $\phi_k$ legs, which are proportional to
\begin{align}
{\rm Tr}(m^e\phi \phi) &= m_1^e\phi_{11}\phi_{11}+m_2^e\phi_{22}\phi_{22}\\
&=m_2^e(\phi_{11}^2+\phi_{22}^2)+(m_1^e-m_2^e)\phi_{11}^2.
\label{eq:2loop}
\end{align}
Here, $\phi={\rm diag}(\phi_{11},\phi_{22})$ with $\phi_{11}=v\cos^2(a/v)$ is obtained from Eq.~\eqref{eq:toystring}, where we promoted the $\theta$-angle into the dynamical Goldstone field $a$ via $\theta/2\equiv a/v$. The first term in Eq.~\eqref{eq:2loop} is proportional to ${\rm Tr}(\phi^2)$ and does not contribute to the Goldstone boson masses. Only the second term, which is proportional to the \textit{difference} between the electron and muon masses, contributes to the Goldstone boson masses. Thus, we observe that $g_W\neq 0$ is not the only requirement to generate nonzero Goldstone boson masses, but we also require $m_1^e\neq m_2^e$. This is analogous to the well-known Glashow–Iliopoulos–Maiani (GIM) mechanism~\cite{Glashow1970} in the weak interactions. 

Using Eq.~\eqref{eq:2loop} and the definitions below, we can now write down the effective Goldstone Lagrangian
\begin{equation}
\mathcal{L}_{{\rm toy,}\textit{a}} = \partial_\mu a\partial^\mu a - \frac{1}{16\pi^2}g^2 g_W^2 v^2 (\delta m_\alpha^e)^2 \cos^2\left(\frac{a}{v}\right),
\label{eq:toyGoldstone}
\end{equation}
where $\delta m_\alpha^e=m_1^e- m_2^e$. Thus, we find a Goldstone boson mass that scales \textit{linearly} with the coupling constant $g_W$,
\begin{equation}
m_a = \frac{1}{4\pi}g g_W (\delta m_\alpha^e).
\label{eq:toymass}
\end{equation}
Note that this nonzero Goldstone boson mass is a prerequisite for the cosmic string~\eqref{eq:toystring} to have a nontrivial impact on the Goldstone Lagrangian~\eqref{eq:toyGoldstone}. If we had no Goldstone boson mass, the second term in Eq.~\eqref{eq:toyGoldstone} would vanish, and the effect of the cosmic string could be completely absorbed into the kinetic term,
\begin{align}
{\rm Tr} \left[\partial_\mu \omega_3^T(\theta)v\,\omega_3(\theta) \partial^\mu \omega_3^T(\theta)v\,\omega_3(\theta)\right]= \partial_\mu a\partial^\mu a.
\end{align}

Up to now, we assumed that the electron and muon in our toy model are Majorana fermions. This allowed for two-loop diagrams with two external $\phi_k$ legs, because the Majorana fermion loops only required a single Majorana mass insertion. Similarly, the electron and muon species in our toy model generated an additional one-loop mass contribution for the Majorana neutrinos.

In the Standard Model, the electron and muon species are Dirac fermions, so their one-loop contribution to the Majorana neutrino mass is forbidden by lepton number conservation. Likewise, the two-loop diagrams yielding Eq.~\eqref{eq:2loop} now require a \textit{double} Dirac mass insertion. These diagrams only renormalize the kinetic term of the neutrino and therefore are unobservable. Similarly, one could naively write down one-loop contributions to the kaon masses~\cite{GellMann:1955jx}, but these contributions would just renormalize the kinetic term. Thus, unlike in Eq.~\eqref{eq:toymass}, both the kaon masses and the $\phi_k$ boson masses do not scale \textit{linearly} with $g_W$ but quadratically instead. 

In the Standard Model, the lowest-order diagram that contributes to the $\phi_k$ boson masses is generated through the four-fermion interaction $G_F(\bar{\nu}\nu)(\bar{e}e)$, which yields a one-loop diagram with four external neutrino lines (corresponding to two external $\phi_k$ lines) and a charged-fermion loop. The resulting four-neutrino interaction is given by 
\begin{equation}
\mathcal{L}_{4\nu}= \frac{1}{16\pi^2}G_F^2(\delta m_{l})^2(\bar{\nu}\nu\bar{\nu}\nu),
\label{eq:4nu}
\end{equation}
where $\delta m_{l}=m_{l_1}-m_{l_2}$, just as in Eq.~\eqref{eq:toyGoldstone}. Due to the large charged-lepton mass differences, $m_\tau\gg m_\mu\gg m_e$, we will in the following use $\delta m_{l}\approx m_l$ with $m_l$ being either $m_\tau$ or $m_\mu$ for the different $\phi_k$ bosons. 

In our neutrino mass model, the neutrino condensate is given by $\langle \bar{\nu}\nu\rangle=\Lambda_G^3$, such that the four-neutrino interaction $(\bar{\nu}\nu\bar{\nu}\nu)$ in Eq.~\eqref{eq:4nu} gets replaced by $\Lambda_G^6(\phi_k/\Lambda_G)^2$. Thus, the effective Goldstone Lagrangian has the following mass term:
\begin{equation}
\mathcal{L}_{{\rm mass},\phi_k} = \frac{1}{16\pi^2}G_F^2m_l^2\Lambda_G^4\,\phi_k^2.
\end{equation}
The resulting $\phi_k$ boson masses scale quadratically in the weak coupling, linearly in the charged-lepton mass difference, and quadratically in the infrared graviational scale,
\begin{equation}
m_{\phi_k}=\frac{1}{4\pi} G_F m_l\, \Lambda_G^2.
\label{eq:finalmass}
\end{equation}

\section{Cosmological evolution of string network in Majorana neutrino case}\label{app:strings}

In this section, we study the cosmological evolution of the string network that forms in the Majorana neutrino case due to the breaking $SU(3)_L\rightarrow Z_2\times Z_2$, see Eq.~(18) in the main text. This string network interacts with the $\phi_k$ bosons that emerge from the late cosmological neutrino annihilation, as we discussed in the section on the relic neutrino phase transition. 

To begin with, we determine how the characteristic length scale $\xi$ of the string 
network grows in response to the effects of acceleration and friction. Due to its tension 
$\mu\sim\Lambda_G^2$, a string with curvature radius $r$ (assumed to be comparable
to the string separation scale $\xi$), experiences a transverse 
accelerating force per unit length,
\begin{equation}
F_{\rm acc} = \frac{\Lambda_G^2}{r},
\label{eq:Facc}
\end{equation}
which competes with the damping force per unit length,
\begin{equation}
F_{\rm damp} = \rho_{\phi_k} \sigma_T v.
\label{eq:Fdamp}
\end{equation}
Here, $\rho_{\phi_k}\sim T_\nu^4$ is the energy density of the neutrino-composite $\phi_k$ bosons, 
which is roughly given by the temperature $T_\nu$ of the cosmic neutrinos in the absence of neutrino 
annihilation. 
Moreover, $v$ is the string velocity and $\sigma_T$ is the thermal scattering cross section per unit length of the strings and the $\phi_k$ bosons, which is given by
\begin{equation}
\frac{\sigma_T}{\Lambda_G} = \alpha_{\rm string-\phi_k}^2\frac{1}{\lambda^2\Lambda_G^2}=\left(\frac{T_\nu^2}{\Lambda_G^2}\right)^2\frac{1}{\lambda^2\Lambda_G^2}=\frac{T_\nu^4}{\lambda^2\Lambda_G^6}
\label{eq:sigmaT}
\end{equation}
due to the derivative coupling of the Goldstone bosons $\phi_k$ to the scalar string field $\Phi_{ij}$. The size of the string core is $1/\lambda\Lambda_g$, where $\lambda$ is the self-coupling of $\Phi_{ij}$.

Due to the large damping force, the evolution of the strings is friction dominated. When comparing Eqs.~\eqref{eq:Facc} and \eqref{eq:Fdamp}, we find that the maximal string velocity reads
\begin{equation}
v_{\rm max} = \frac{\Lambda_G^2}{r\rho_{\phi_k}\sigma_T} = \frac{\lambda^2\Lambda_G^7}{rT_\nu^8} \equiv \frac{l_F}{r},
\label{eq:vmax}
\end{equation}
where $l_F$ is the length scale associated with friction.
The expected time scale $\tau$ on which $\xi\sim r$ increases is
\begin{equation}
\frac{\dot{\xi}}{\xi}=\frac{1}{\tau}=\frac{v_{\rm max}}{\xi}=\frac{l_F}{\xi^2}=\frac{\lambda^2\Lambda_G^7}{\xi^2T_\nu^8}.
\end{equation}
When comparing this time scale with the Hubble time $t$ and going from comoving to proper distances, we obtain for the final characteristic length scale of the strings
\begin{equation}
\xi =\sqrt{\frac{\lambda^2\Lambda_G^7 t}{a_G^2 T_\nu^8}}.
\label{eq:xi}
\end{equation}

\section{Annihilation of skyrmions, monopoles, and string-wall network in Dirac neutrino case}\label{app:annihilation}

In this section, we discuss the formation and evolution of further topological defects that are not expected to have any observational consequences. In particular, we study the skyrmions, monopoles, and string-wall network that arise in the case of Dirac neutrinos. Note that a similar string-wall network also arises in the Majorana case, due to the breaking $U(1)_L\to Z_2$ in Eq.~(10) of the main text.

\textit{Skyrmions}---For Dirac neutrinos, the breaking of the $SU(3)_A$ symmetry in Eq.~(15) of the main text gives rise to textures, 
\begin{align}
\pi_3(SU(3))=Z.
\end{align}
Unless stabilized by the Skyrme term~\cite{Skyrme1962}, the textures are unstable, i.e.\ the topological texture 
knots quickly unwind with a rate $dn/dt\sim \Lambda_G^4$ and dissipate into Goldstone bosons \cite{Turok1989}. 
In the neutrino mass model, the textures are stabilized by the Skyrme term~\cite{Dvali2016a}, just as in the QCD baryon case~\cite{Adkins1983}. The skyrmions are spin $1/2$ states because they are bound states of $N_F=3$ neutrinos. 

After formation, the skyrmions and anti-skyrmions annihilate into $\phi_k$ bosons with an annihilation rate
of~\cite{Gillioz2010}
\begin{align}
\Gamma_S&=\langle \sigma_S v_S\rangle \, n_{\rm eq,S}\nonumber\\
&=\frac{\pi}{\Lambda_G^2}\left(\frac{3T_\nu}{\Lambda_G}\right)^{1/2}\left(\frac{\Lambda_G T_\nu}{2\pi}\right)^{3/2}e^{-\Lambda_G /T_\nu},
\end{align}
where $\sigma_S=\pi/m_S^2\sim\pi/\Lambda_G^2$ is the geometrical cross section, $m_S\sim\Lambda_G$ is the skyrmion mass, $v_S$ is the thermal skyrmion velocity, and $n_{\rm eq,S}$ is the skyrmion equilibrium density. This annihilation is less efficient than, e.g., for global monopoles, due to negligible long-range forces between skyrmions and antiskyrmions \cite{Jackson1985}. Nevertheless, the skyrmions quickly annihilate into the massless $\phi_k$ bosons within timescales of $\Gamma_S^{-1} = 10^{-15}~{\rm s}$ for the exemplary values of $\Lambda_G= T_\nu= $~0.1~eV, and only a negligibly small freeze-out density remains.

\textit{Global monopoles}---The $SU(3)_V$ symmetry breaking in Eq.~(17) of the main text gives rise to global monopoles carrying two $U(1)$ charges,
\begin{align}
\begin{split}
\pi_2(SU(3)_V/ [U(1)_3\times U(1)_8])=Z\times Z.
\end{split}
\label{eq:pi_2}
\end{align}
These monopoles have energies of the order $m_M\sim\Lambda_G^2 d(t)$, where $d(t)$ is the time-dependent distance to the 
nearest antimonopole (with $d(0)\sim\Lambda_G^{-1}$).
The attractive force between global monopoles and anti-monopoles is proportional to 
$\Lambda_G^2$ and independent of the distance between them~\cite{Barriola1989}. Thus, 
the monopole-antimonopole separation decreases at relativistic speeds and the (anti)monopoles 
efficiently annihilate into $\phi_k$ bosons~\cite{Martins:2008ks}. 

\textit{String-wall network}---The anomalous $U(1)_A$ symmetry in Eq.~(15) of the main text has an anomaly-free $Z_6$ subgroup, similar to axion scenarios~\cite{Peccei1977,Weinberg1977, Wilczek1978,Kim1979, Shifman1979, Zhitnitsky1980,Dine1981}. 
Thus, the spontaneous symmetry breaking of this anomalous $U(1)_{A}$ symmetry gives rise to strings and DWs~\cite{Sikivie1982,Vilenkin1982},
\begin{equation}
\pi_1(U(1)_{A}/Z_6)=Z,\quad \pi_0(Z_6/Z_2)=Z_3,
\label{eq:pi_10}
\end{equation}
For $N_F=3$ neutrino species, three DWs attach to each string, and a hybrid string-wall network forms.

The thickness of the string core is given by $\delta_S\sim\Lambda_G^{-1}$ and the mass per unit area of the DWs is $\sigma_W\sim\Lambda_G^3/N_F$.
Note that the resulting DW width is much smaller than the width of the $Z_2\times Z_2$ DWs in Eq.~(22) of the main text, because here the dominant contribution to the explicit symmetry breaking is the anomaly, not the weak effects given by Eq.~(4) of the main text. Therefore, at the time of formation, the DW and string energy densities are comparable. The tension of the strings efficiently straightens them out, and the string energy density soon becomes negligible compared 
to the DW energy density. Thus, the DWs determine the dynamics of the string-wall network. 
Numerical simulations~\cite{Vilenkin1982,Ryden1990,Leite2011} showed that the string-wall energy 
density eventually approaches a scaling solution, 
\begin{equation}\label{eq:rho_WS1}
\rho_{\rm SW}(t)\sim \sigma_W t^{-1}\sim \Lambda_G^3 t^{-1},
\end{equation}
such that $\mathcal{O}(10)$ DWs remain per Hubble volume. 

This DW density is too low to yield any observational consequences, similar to the skyrmion and monopole densities discussed above. The reason for the low density of these DWs compared to the high densities of the DWs discussed in the main text is that the current DWs do not interact with the ambient cosmological medium. Due to the axionic nature of the current DWs, the $\phi_k$ bosons pass through them without reflection. Thus, these axionic DWs are not slowed down by friction and their annihilation is very efficient~\cite{Vilenkin2000}.

In the neutrino mass model~\cite{Dvali2016a}, the $\eta_\nu$ bosons resulting from the string-wall annihilation further decay into the $\phi_k$ bosons and thus contribute to dark radiation in the late Universe. In contrast, the gravitational axion model proposed in~\cite{Dvali2016b} only requires the $\eta_\nu$ boson, which is stable and thus may contribute to the late dark matter abundance. The precise density of these dark bosons strongly depends on the free parameters of the models, such as the scale $\Lambda_G$ and the phase transition time. 

As a general remark regarding the energy density of the dark radiation in the Dirac case, we wish to emphasize that this density also depends on the nature of the phase transition. If the transition is a smooth, second-order transition, the resulting energy density of the dark radiation is given by the energy density of the neutrinos at the phase transition time. If the phase transition is a supercooled, first-order transition, the energy density of the dark radiation could be much larger, as discussed briefly in the main text and in more detail in Ref.~\cite{Lorenz:2018fzb}.

\section{Non-topological domain walls}

In this section, we briefly discuss the non-topological DWs that arise when the Universe falls into different vacua with different neutrino mass matrices after the phase transition. The effective neutrino potential~\cite{Dvali2016a,Dvali2016b} 
\begin{equation}
V(\hat{X}) =  \sum_n \frac{1}{n} c_{2n} {\rm Tr}[(\hat{X}^+\hat{X})^n],
\label{eq:effpotential}
\end{equation}
allows for a large variety of neutrino flavor symmetry breaking patterns, parametrized by unknown coefficients $c_{2n}$ that multiply generic invariants ${\rm Tr}[(\hat{X}^+\hat{X})^n]$ of the neutrino condensate order parameter $ \hat{X}_{\alpha_L}^{\alpha_R}\equiv\langle \bar{\nu}_{\alpha_L} \nu_{\alpha_R} \rangle$. Here, $\alpha_L = 1,2,3$ and $\alpha_R = 1,2,3$ are left- and right-handed flavor indices, respectively. The effective potential in Eq.~\eqref{eq:effpotential} allows for a highly nontrivial vacuum structure. Thus, when the Universe falls into different vacua after the transition, these different vacua on either side of the resulting DWs are not exactly degenerate.

The evolution of the resulting network of non-topological DWs depends on the pressure difference $\delta p$ 
across the DWs as compared to the viscosity $\eta$ of the DWs due interactions with the ambient cosmological medium. Depending on the unknown coefficients $c_{2n}$ in Eq.~\eqref{eq:effpotential}, the pressure difference $\delta p$ can range between $\delta p=0$ and $\delta p=m_{\nu_i}\lesssim 0.8$~eV~\cite{Aker:2021gma}, while the viscosity is determined by the density $\rho_{\phi_k}\sim T_\nu^4\sim (0.1~{\rm meV})^4$ of the $\phi_k$ bosons that scatter off the DWs (see Eq.~\eqref{eq:Fdamp} for more details). 
Thus, for most of the DWs, the pressure difference is expected to be larger than the frictional force, such that the DWs quickly annihilate and dilute. The observational consequences of these non-topological DWs are therefore similar to the ones discussed at the end of the previous section.

\bibliographystyle{apsrev4-2}
\interlinepenalty=10000
\bibliography{Bibliography}

\end{document}